\documentclass[article,aps,prl,twocolumn,floatfix,superscriptaddress]{revtex4-1}
\usepackage{graphicx}
\usepackage{dcolumn}
\usepackage{bm}
\usepackage{amsmath}
\usepackage[english]{babel}
\usepackage[utf8]{inputenc}
\usepackage{color}
\usepackage{upgreek}

\usepackage{siunitx}

\newcommand{\blue}[1]{\textcolor{blue}}
\usepackage[T1]{fontenc}
\usepackage[hidelinks]{hyperref}
\hypersetup{
    colorlinks=true,
    citecolor=magenta,
    linkcolor=blue,
    filecolor=magenta,      
    urlcolor=magenta,
    }

\begin{document}

\title{Confinement drives valley splitting above 4K in buried silicon quantum wells}


\author{Davide Degli Esposti}
\email{d.degliesposti@tudelft.nl}
\affiliation{QuTech and Kavli Institute of Nanoscience, Delft University of Technology, Lorentzweg 1, 2628 CJ Delft, The Netherlands}
\author{Emma C. Brann}
\affiliation{Department of Physics, University of Wisconsin-Madison, Madison, Wisconsin, 53706, United States}
\author{Asser Elsayed}
\affiliation{QuTech and Kavli Institute of Nanoscience, Delft University of Technology, Lorentzweg 1, 2628 CJ Delft, The Netherlands}
\author{Davide Costa}
\affiliation{QuTech and Kavli Institute of Nanoscience, Delft University of Technology, Lorentzweg 1, 2628 CJ Delft, The Netherlands}
\author{Mark Friesen}
\affiliation{Department of Physics, University of Wisconsin-Madison, Madison, Wisconsin, 53706, United States}
\author{Giordano Scappucci}
\email{g.scappucci@tudelft.nl}
\affiliation{QuTech and Kavli Institute of Nanoscience, Delft University of Technology, Lorentzweg 1, 2628 CJ Delft, The Netherlands}

\date{\today}

\begin{abstract}

Controlling the energy scales of a quantum system is essential for defining robust qubits. 
In silicon spin qubits, the nearly degenerate conduction-band valleys create a leakage channel from the single-spin computational basis, posing a challenge to scaling and to shuttling-based architectures.
Here, we measure the relevant energy scales of single-electron spin qubits in buried silicon quantum wells co-designed for low disorder and high valley splitting. 
Across a linear array of four quantum dots with an average orbital energy of 2.4(2) meV, we report an average single-electron valley splitting of 0.40(6) meV and an average two-electron singlet-triplet splitting of 0.24(7) meV.
In three dots, we observe a strong correlation between valley splitting and orbital energy, with an average linear coefficient of $\approx 0.22$ (meV/meV), demonstrating that electrostatic confinement can increase the valley splitting by several hundred microelectronvolts.
In contrast, the remaining dot exhibits the highest valley splitting of 0.76(2) meV and low correlation, suggesting excellent characteristics for spin-qubit operation.
Our findings demonstrate that strong confinement can be exploited in buried quantum wells to effectively enhance the valley splitting, thereby establishing a viable path toward the realization of shuttling and sparse-occupation-based architectures in low-disorder heterostructures.

\end{abstract}

\maketitle

\section{Introduction}

Gate-defined quantum dots in $^{28}$Si/SiGe heterostructures are a promising platform for building large-scale quantum processors~\cite{burkard2023semiconductor,stano2022review}. 
Indeed, the key building blocks of a large quantum processor, such as high-fidelity single and two-qubit operation~\cite{xue_quantum_2022,wu2025simultaneous}, coherent shuttling and teleportation to distribute quantum information~\cite{de2025high,matsumoto2025two,yoneda2021coherent,fujita2017coherent}, prolonged digital quantum circuit operation~\cite{fernandez2026running,capannelli2025tracking}, and first error correction schemes~\cite{undseth2026weight}, have been demonstrated.
Moreover, the advanced semiconductor industry proved to be able to manufacture uniform and reliable spin qubit devices ~\cite{marcks2025valley,mkadzik2025operating,george202412,steinacker2025industry,fattal2025radio,neyens2024probing,koch2025industrial,zwerver2022qubits,neyens2024probing,steinacker2025industry}.
The buried isotopically purified quantum well ensures a quiet environment with low electrical and magnetic noise~\cite{zwanenburg_silicon_2013,neyens2024probing}. 
Low disorder and the absence of decoherence hotspots will be a desirable requirement for scaling to large arrays and developing shuttling-based or sparse-array architectures ~\cite{undseth2026weight,siegel2025snakes,krzywda2026coherence,li2025trilinear,kunne2024spinbus, volmer2024mapping,losert2024strategies}.  
In this context, the uncontrollable and frequent occurrence of degenerate valley states near the computational single-spin basis remains a major obstacle for scaling to large arrays, connecting distant registers via shuttling links, and operating at high temperature ~\cite{kawakami_electrical_2014,philips_universal_2022,degli2024low,borselli_measurement_2011,shi_tunable_2011,mi_high-resolution_2017,hollmann_large_2020,mcjunkin_valley_2021,chen_detuning_2021,dodson_how_2022,denisov2022microwave}. 



In silicon metal-oxide-semiconductor (SiMOS), the strong electrostatic confinement at the semiconductor-dielectric interface has enabled gate-tuneable valley splitting up to $\approx 0.8$~meV~\cite{yang_spin-valley_2013} and high-fidelity qubit operation at high temperature~\cite{nickl2025eight,huang2024high}.
On the contrary, Si/SiGe heterostructures are characterized by a shallower confinement due to the larger distance between the gates and the buried quantum well set by the epitaxial SiGe barrier.
Multiple experimental and theoretical works investigating the microscopic origin of the valley separation in Si/SiGe demonstrate that alloy disorder, intended as the random distribution of Ge atoms in the SiGe alloy, constitutes the dominant factor driving the valley splitting in these heterostructures~\textmu eV~\cite{paquelet_wuetz_atomic_2022,losert_practical_2023,Fiesen2007VS_theory, culcer2010interface,hosseinkhani2020electromagnetic,salamone2025valley}. 
As a result, different strategies have been proposed and are experimentally investigated to engineer the disorder potential and systematically enhance the valley splitting~\cite{losert_practical_2023}, including the use of unconventional heterostructures that incorporate Ge into the interior~\cite{mcjunkin_valley_2021,feng_enhanced_2022,paquelet_wuetz_atomic_2022,mcjunkin_sige_2022,stehouwer2025engineering,thayil2025optimization} or the boundary of the quantum
well~\cite{zhang_genetic_2013,neyens_critical_2018,wang_origin_2022}.
However, all of these strategies aim to increase alloy scattering within or near the quantum well, thereby compromising the low-disorder environment characteristic of these heterostructures, and have so far resulted in only nondeterministic improvements~\cite{george202412}.
Recently, low disorder and high valley splitting have been demonstrated in conventional $^{28}$Si/SiGe heterostructures~\cite{degli2024low}. 
Such heterostructures enabled more complex devices to demonstrate high-fidelity spin shuttling \cite{de2025high} and weight-four parity stabilizers~\cite{undseth2026weight}, suggesting a possible co-integration of these characteristics.

Here, we advance our investigations of the same low-disorder shallow quantum wells and report spectroscopic measurements of single- and two-electron states in tightly confined quantum dot devices embedded in these heterostructures.
We measure all the relevant energy scales for defining single-electron spin qubits, such as the charging energy ($E_C$), the orbital energy ($E_O$), the single-electron valley splitting ($E_V$), and the two-electron singlet-triplet splitting via magnetospectroscopy ($E_{ST}$) and Pauli-Spin-Blockade spectroscopy ($E_{PSB}$). 
By using gates to tune the confinement, we find a strong correlation between $E_O$ and $E_V$ over multiple quantum dots, boosting $E_V$ above $0.5$~meV.
This correlation breaks down in one quantum dot, which features both the largest $E_O$ of  $2.77(4)$~meV and $E_V$  valley splitting of $0.76(2)$~meV.
We compare our measurement with current theories of valley splitting in the alloy-disorder-dominated regime and reveal that confinement might be surprisingly efficient in these heterostructures at enhancing the single-electron valley splitting. 
Despite the tight confinement, we find a substantial reduction of the two-electron singlet-triplet splitting in the same dots compared to the respective valley splitting.   
This study underscores the importance of characterizing energy scales beyond single-electron states and provides a viable path to co-designing quantum dots and buried quantum wells with low disorder and consistently large valley splitting. 

\section{Results}

\subsection{Heterostructure and device characterization}

\begin{figure*}[t]
	\includegraphics[width=165mm]{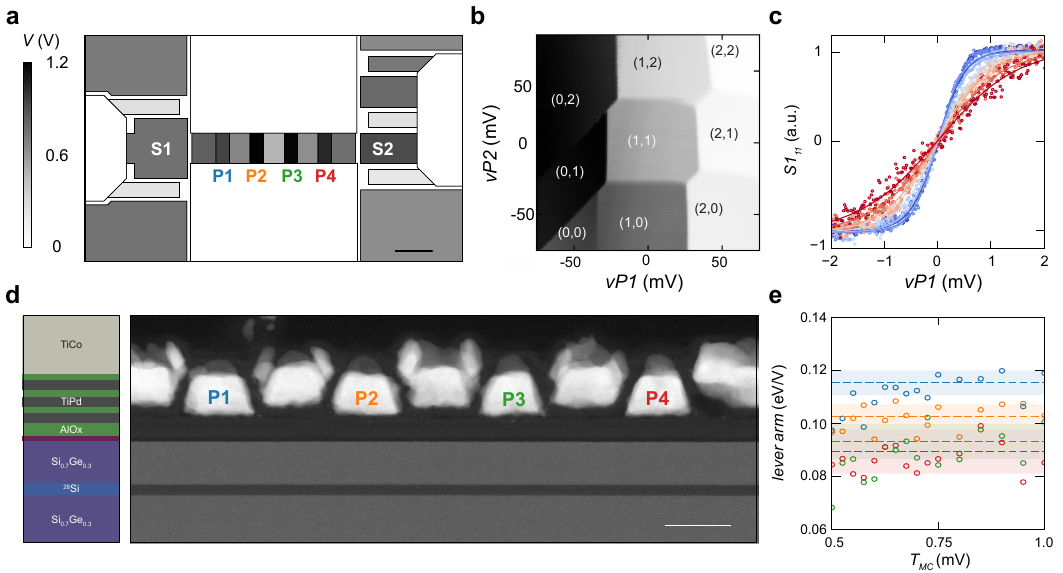}%
\caption{\textbf{Heterostructure and quantum dot gate stack} 
\textbf{a} Quantum dot gate layout and voltage configuration used to achieve single-electron occupancy in the four-dot array.  
The top-right sensor is fully accumulated and used as an extension of the accumulation gate.
The two sensors (S1 and S2) are used to sense the charge state of the four quantum dots (P1 to P4).
The black line corresponds to 100 nm.
\textbf{b} Double-dot charge stability diagram centered in the DC voltage configuration of \textbf{a} for the virtualized plunger gates $vP_1$ and $vP_2$. 
\textbf{c} First electron polarization line for virtual plunger vP1 measured as a function of mixing chamber temperature for temperatures between 10 mK (dark blue) and 1 K (dark red).
The solid line is a fit to the theoretical model used to extract the lever arm.
\textbf{d} Atomic-resolution high-angle annular dark-field (HAADF) (Z-contrast) scanning transmission electron microscopy (STEM) image of the overlapping gate and heterostructure stack taken along the linear quantum dot array.
The gates corresponding to the quantum dot plungers are labeled P1-P4.
The sketch on the left side highlights the layer order forming the heterostructure and gate stack.
The white line corresponds to 50 nm.
\textbf{e} Average lever arm (dashed line) and standard deviation (shaded area) for the four virtual plungers vP1 (blue), vP2 (orange), vP3 (green), and vP4 (red) obtained from the fitting of the first electron polarization line to the Fermi distribution for temperatures of the mixing chamber between 0.5 and 1 K. }
\label{fig:1}
\end{figure*}

We fabricated quantum dot spin qubit devices on top of $^{28}$Si/SiGe heterostructures grown by reduced-pressure chemical vapor deposition (RP-CVD) on 100 mm Si(001) substrates~\cite{degli_esposti_wafer-scale_2022,degli2024low} using lift-off processes and overlapping gates~\cite{philips_universal_2022} (see Methods).
We use shallow quantum wells buried $30$ nm beneath the surface and thin high-k dielectrics in the gate stack to maximize the capacitive coupling between the charges in the quantum well and the topmost electrodes.   
From the measurement of the two-dimensional electron gas (2DEG), we extract an effective capacitance of $C_{2DEG} \simeq 210$ nF/cm$^2$\cite{degli_esposti_wafer-scale_2022} at the first gate layer. 
Fig.\ref{fig:1}(a) shows the device layout and the voltage configuration used to achieve the single-electron occupancy in the quantum dot array. 
The array comprises four quantum dots (P1-P4), a single charge sensor on the left side (S1), and a double charge sensor on the right side (S2). 
We design these devices with a tight gate pitch of $90$~nm, a nominal single-dot plunger width of $40$~nm, and strong in-plane asymmetry to lift the in-plane orbital states.
We operate the right sensor, S2, as a single-charge sensor defined by the bottom plunger.
We connect the two sensing dots to two off-chip high-inductance superconducting resonators to perform fast readout via frequency-multiplexed RF-reflectometry (see Supplementary Information) and perform all experiments at a constant mixing chamber temperature of $150$ mK to minimize drift and heating effects~\cite{undseth_hotter_2023}.

We populate the array in the (1,1,1,1) electron configuration and estimate the cross-capacitance coupling between gates from the virtual-gate matrix.
In the Supplementary Information, we report the virtualization matrix for the main device and for a lithographically identical device to validate the reproducibility of the gate layout and cross-capacitive coupling.
Fig.\ref{fig:1}(b) shows a charge stability diagram centered in the single-electron configuration for the virtualized plungers vP1 and vP2.
The symmetry of the devices is confirmed by the high-resolution scanning transmission electron micrograph (STEM) image in Fig.\ref{fig:1}(c) of the plungers and barriers along the array direction.
From the STEM image, we measure a plunger (barrier) width of $40$~nm ($50$~nm) and confirm it by scanning electron microscopy and atomic force microscopy of lithographically identical devices (see Supplementary Information).

We continue to characterize the quantum dot device and measure the lever arm of each virtualized plunger gate via temperature-activation broadening of the first electron addition line~\cite{de2025high}. 
We maintain a low tunnel coupling between the dots and measure the first-electron-addition line at temperatures well above this energy scale to ensure that the broadening is temperature-limited. 
Fig.\ref{fig:1}(c) shows the first polarization line as a function of vP$1$ for temperatures of the mixing chamber between $10$~mK and $1$~K. 
We fit the addition line with the phenomenological relation $A + B + C\tanh{\alpha(V - V_0)\beta}$ where $A$, $B$, and $\alpha$ are fitting parameters and $\beta = \frac{e}{2 k_B T_{e}}$  with $k_B$ the Boltzmann constant and $T_e$ is the electron temperature that we approximate with the temperature of the mixing chamber in this regime. 
Fig.\ref{fig:1}(e) shows the results of the fit at temperatures between $0.5$ and $1.0$~K, where the addition line is mostly temperature broadened. 
The extracted lever arms remain constant in this regime and converge to values between $0.08$ and $0.12$ eV/V. 

\subsection{Single-electron spectroscopy}

\begin{figure}[t]
	\includegraphics[width=86mm]{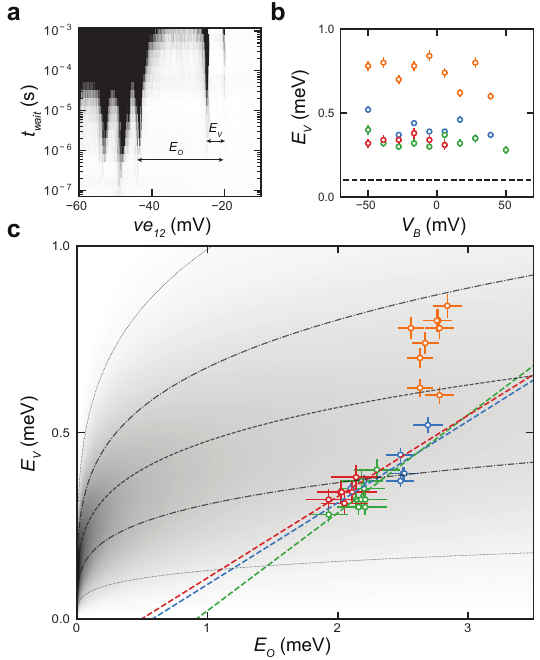}%
\caption{\textbf{Single-electron spectroscopy} 
\textbf{a} Detuned axis pulsed spectroscopy (DAPS) maps of dot P1 obtained by sweeping the relative detuning ($ve_{12}$) for different waiting times $t_{wait}$.
Charge tunneling at a specific detuning is enhanced by the presence of well-defined quantum states, giving rise to a faster relaxation rate.
We extract $E_O$ and $E_V$ from the separation of the resonant peaks and convert them into energy using the specific lever arm for each gate.
\textbf{b} Single-electron valley splitting ($E_V$) extracted from DAPS for the four dots (P1 blue, P2 orange, P3 green, and P4 red) as a function of the voltage applied to the barriers ($V_B$) between the dots. 
The dashed line is set at 0.1 meV, which is typically taken as the minimum required to ensure readout fidelities above 99\%~\cite{paquelet_wuetz_atomic_2022}.
\textbf{c} Cross correlation plot between $E_V$ and $E_O$.
We fit the values from  P1, P3, and P4, showing a high correlation with a linear function of the form $E_v = \alpha_{E_V/E_O} E_O + b$, from which we extract $\alpha_{E_V/E_O} = 0.22(4)$ [eV/eV] and $b = -0.12(8)$ meV. 
The shaded area represents the theoretical probability density distribution for $E_V$ as a function of $E_O$, obtained by fitting the average z-y integral to the experimental data (see Methods). 
The dark solid lines follow the median (dashed), 25th to 75th percentile (dot-dashed), and 5th to 95th percentile (dotted) of the theoretical valley splitting distributions. }
\label{fig:2}
\end{figure}


We start the characterization of the valley and orbital states by performing detuning axis pulsed spectroscopy (DAPS)~\cite{chen_detuning_2021} on the entire array. 
We prepare one electron in one of the dots and track the resonant tunneling to the nearby dot across the (1,0) to (0,1) transition as a function of the relative detuning ($ve_{12}$ and $ve_{43}$) and use the previously measured lever arm to convert into an energy spectrum. 
This technique allows us to extract the energy spectrum in a configuration in which the quantum dot array is strongly decoupled from the sensing dot and the number of electrons in the array is constant. 
Fig.\ref{fig:2}(a) shows the results of this measurement for a single electron tunneling from P1 to P2 as a function of the waiting time ($t_{wait}$) and detuning ($ve_{12}$).  
We recognize two pairs of resonant tunneling states that we attribute to the first valleys and orbital states~\cite{chen_detuning_2021} and use the separations to extract $E_V$ and $E_O$. 

We repeat the measurement for each dot in the array and for different confinement configurations, controlled by the barrier potential ($V_B$) between the dots, to study the effect of electrostatic confinement on $E_V$ and $E_C$. 
Fig.\ref{fig:2}(b) shows the results of this analysis for the measured single-electron valley splitting for the four quantum dot arrays. 
Within the variability expected for the valley splitting~\cite{marcks2025valley}, these quantum dots show consistently large values across a wide range of gate voltages, a necessary condition for high-fidelity operation during long pulse sequences implementing a quantum algorithm. 

At the same time, we extract the first orbital energy from the DAPS spectrum for each confinement configuration.
We plot the measured values of $E_V$ and $E_O$ in Fig.\ref{fig:2}(c) to analyze their correlation.
We quantify the correlation between $E_O$ and $E_V$ using a Pearson correlation test, where $1$ indicates a perfect correlation and $0$ indicates no correlation, and use the corresponding p-value to assess the statistical significance of each test (see Supplementary Table S2). 
For the dots defined by P1, P3, and P4, we find correlation coefficients of $\approx 0.7$, indicating a strong positive correlation between $E_V$ and $E_O$. 
This observation is consistent with the current theories of valley splitting in the random-alloy disorder-dominated regime and is expected for these heterostructures~\cite{losert_practical_2023}. 

To further characterize the relation between $E_O$ and $E_V$, we fit the datasets for P1, P3, and P4 with a linear relation of the form $E_V = \alpha_{E_V/E_O} E_O + b$ (colored dashed lines in Fig\ref{fig:2}(c)).
The wavefunction of the dots with large $E_O$, and therefore stronger electrostatic confinement, is strongly squeezed at the Si/SiGe interface and is expected to sample a larger distribution of Ge atoms in the SiGe spacer, enhancing the alloy-disorder coupling. 
The linear coefficient for each dot quantifies the local alloy-disorder landscape of the specific dot.
We find, for all dots, a steep linear coefficient $\alpha_{E_V/E_O} \approx 0.20 - 0.24$ meV/meV, indicating that electrostatic confinement can be an effective strategy for tuning the valley separation by hundreds of micro-electronvolts.
The linear trend between valley splitting and confinement is also measured in the 2DEG of similar heterostructures as a function of the out-of-plane magnetic field~\cite{stehouwer2025engineering}. 


We report the highest $E_V$ within the array for the dot under P2, with an average $E_{V-P2} = 0.72(1)$ meV and a low correlation coefficient of $\approx 0.3$. 
These values set a new benchmark for single-electron valley splitting in buried quantum wells and are on par with the best values reported in SiMOS devices~\cite{yang_spin-valley_2013}.
The low correlation is generally unexpected for quantum dots defined in low-disorder heterostructures and could have multiple origins. 
On the one hand, the low correlation between $E_O$ and $E_V$ is consistent with a deterministic enhancement of the valley splitting beyond the alloy-disorder-dominated regime. 
On the other hand, it could be explained by the specific quantum dot pinning with a disorder center modifying the confinement and the valley coupling, besides the gate effects, and in line with the current theories~\cite{losert2024strategies}. 
While these measurements do not allow us to discern a specific origin, new techniques are currently being developed to better pinpoint the microscopic origin of valley-splitting~\cite{woods2025statistical}. 

Next, we compare the experimental data with the theoretical predictions for single-electron quantum dots in these heterostructures~\cite{losert_practical_2023}.
The theoretical distribution of the valley splitting depends on the electric field at the quantum-well interface ($E_Z$), the microscopic Ge-alloy disorder, and the orbital energy~\cite{losert_practical_2023}. 
We estimate the electric field ($E_Z \approx 10$ MV/m) from the maximum charge density in the two-dimensional electron gas~\cite{degli_esposti_wafer-scale_2022}.
The alloy disorder enters the theoretical model when integrating over the z-coordinate and can be approximated phenomenologically as a constant that depends on the specific Ge concentration profile of the SiGe/Si/SiGe heterostructure, as determined by STEM (see Supplementary Information).   
Contrary to the experimental results, the calculated valley-splitting distribution (see Supplementary Fig. S12) predicts that most of our values should remain below $0.2$ meV. 

We then fit the disorder parameter to the experimental distribution of $E_V$ to assess if the theoretical model can predict the spread and dependence of valley splitting as a function of orbital energy. 
We plot the distribution's probability density as a function of orbital energy as a shaded area in Fig.\ref{fig:2}(c) and highlight its evolution as a function of the orbital energy by plotting the median (dashed line), 25th to 75th percentile (dot-dashed line), and 5th to 95th percentile (dotted) of the theoretical distributions.
As expected, the spread of the valley-splitting distribution increases with its average and with the orbital energy~\cite{paquelet_wuetz_atomic_2022} and matches the observed spread in the experimental data. 

\subsection{Two-electron spectroscopy}

\begin{figure}[t]
	\includegraphics[width=86mm]{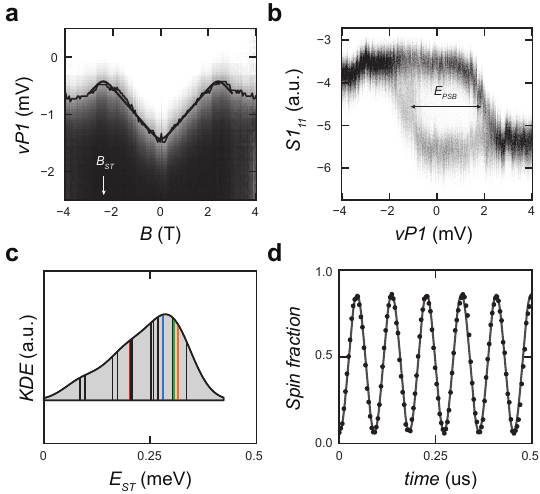}%
\caption{\textbf{Two-electron spectroscopy.} 
\textbf{a} Magnetospectroscopy of the second-electron addition line under vP1 as a function of the external in-plane magnetic field $B$.
The thick dark line is a fit of the transition line to the theoretical model from which we extract the position of the kink at $B_{ST} = 2.4(1)$ T, corresponding to a singlet-triplet splitting of $E_{ST} = 0.28(1)$ meV, and a lever arm of $\alpha = 0.11$ eV/V in agreement with the previous measurement via temperature activation.
\textbf{b} Histogram of the reflected signal amplitude from the sensing dot S1 for 50\,000 single-shot measurements as a function of vP1 across the (1,1)-(0,2) anticrossing from which we extract the two-electron singlet-triplet splitting from Pauli-Spin-Blockade spectroscopy ($E_{PSB}$). 
\textbf{c} Kernel Density Estimation (KDE) of the probability density for the singlet-triplet splitting extracted from 17 lithographically identical quantum dots from four different devices and two different wafers (9 values from Ref.\cite{degli2024low}). 
The solid lines within the distribution represent the single instances of $E_{ST}$ for each dot; the colored lines correspond to the four dots (P1-P4). 
From the distribution, we extract a mean singlet-triplet splitting of $\bar{E}_{ST} = 0.24(7)$ (meV). 
\textbf{d} Spin fraction for the double dot under P1 and P2 as a function of the waiting time at the symmetry point of the (1,1) charge state, showing coherent oscillation between Singlet and Triplet states mediated by the exchange and used to validate the spin selectivity of the readout scheme before Pauli-Spin-Blockade spectroscopy.}
\label{Fig:3}
\end{figure}

We conclude the characterization by measuring two-electron singlet-triplet energy splitting via magneto-spectroscopy ($E_{ST}$) and Pauli-Spin-Blockade spectroscopy ($E_{PSB}$). 
$E_{ST}$ is the relevant energy scale for the spin-to-charge conversion readout using the Pauli exclusion principle~\cite{philips_universal_2022,borselli_pauli_2011} and therefore sets an ultimate limit to the possible readout fidelity and operation temperature of electron spin qubits in Si/SiGe. 

Fig.\ref{Fig:3}(a) shows a typical magnetospectroscopy map across the $1$e to $2$e transition under the virtual plunger vP$1$ as a function of the in-plane magnetic field (B). 
The thick black line is a fit of the transition line to the theoretical model (see Methods) from which we estimate the singlet-triplet splitting $E_{ST} = g \mu_B B_{ST}$, where $g=2$ is the electron g-factor, $\mu_B$ is the Bohr magneton, and $B_{ST}$ corresponds to the magnetic field at which the final two electron state transition from a singlet ($S_0$) to a triplet ($T_-$) spin state, \emph{i.e.,} the position of the kink in Fig.\ref{Fig:3}(a). 
For the external dots P1 and P4, we perform magnetospectroscopy of the addition line from the reservoir (1e $\rightarrow$ 2e) and the interdot transition ((1,1) $\rightarrow$ (2,0)), and find comparable values of $E_{ST}$.
From the fit to the magnetospectroscopy trace, we also extract the lever arm of each virtual plunger gate in agreement with the values previously measured via temperature activation. 

We collect singlet-triplet energy splitting from 17 lithographically identical quantum dots from four different devices and two nominally identical wafers (9 values from Ref.\cite{degli2024low}) and plot in Fig.\ref{Fig:3}(c) the Kernel Density Estimation (KDE) of the probability density for $E_{ST}$. 
The black lines within the distribution represent the individual instances of singlet-triplet splitting, while the colored lines highlight the instances from the four quantum dots P1 to P4. 
Contrary to the Rayleigh or Rice-like distribution expected for the single-electron valley splitting in the alloy disorder-dominated regime, the measured distribution for $E_{ST}$ shows a bell-like shape with an average singlet-triplet splitting of $E_{ST} = 0.24(6)$ meV.
Notably, in all dots, $E_{ST}$ is substantially lower than the measured $E_V$. 
This suggests that the formation of Wigner molecule states~\cite{abadillo2021two,ercan2021strong,corrigan2021coherent} -- in which strong Coulomb repulsion spatially localizes the two electrons -- limits the singlet-triplet splitting rather than single-electron valley splitting. 

Finally, we corroborate the characterization of two-electron states by performing Pauli-Spin-Blockade spectroscopy across the (1,1) $\rightarrow$ (2,0) transition. 
In this charge configuration, the singlet-triplet splitting is limited by the valley energies and should provide a precise hallmark in the case of low occurrences of valley splitting~\cite{philips_universal_2022}. 
We perform initialization via readout and post-selection on shots with the same initial state. 
By scanning the interdot transition as a function of the plunger gate, the sensor signal splits, and a window opens in which singlet and triplet states are selectively associated with different charge states, known as the PSB window. 
We report the results in Fig.\ref{Fig:3}(b). 
We use the lever arm to convert the width of the PSB window and extract an estimation for the singlet-triplet splitting of $E_{PSB} \approx 0.42(2)$ meV.
We repeat the measurement for the four quantum dots and measure coherent oscillations of the spin fraction after each sequence, as in Fig.\ref{Fig:3}(d), to validate the spin-selectivity of the readout scheme (see Supplementary Information).  

We find that the two-electron $E_{PSB}$ energy splittings are in quantitative agreement with magnetospectroscopy measurements for the internal dots P2 and P3 and are lower than the single-electron $E_V$ values for the respective dots, similarly to what was reported in Ref.\cite{abraham2026digitally}. 
For the external dots, P1 and P4, we find a larger $E_{PSB}$ than $E_{ST}$.  
This can be explained by the different tuning conditions between the two measurement regimes. 
We perform PSB spectroscopy in the so-called isolated mode, \emph{i.e.}, under low tunnel coupling between the sensors and the external dots. 
As a result, the confinement of the external dots is tighter during PSB spectroscopy, leading to higher singlet-triplet energy levels, consistent with the previous interpretation that $E_ST$ is limited by the confinement and the formation of Wigner-molecule states.

\subsection{Discussion and Conclusion}

\begin{figure}[t]
	\includegraphics[width=80mm]{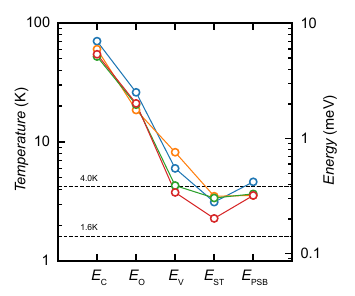}%
\caption{\textbf{Single-electron and two-electron energy states in silicon quantum dots.} 
We report the charging energy ($E_C$), single electron orbital energy ($E_O$), single electron valley splitting ($E_V$), two-electrons singlet-triplet splitting measured via magnetospectroscopy ($E_{ST}$) and Pauli-Spin-Blockade spectroscopy ($E_{PSB}$) for all the four dots (P1 blue, P2 orange, P3 green, and P4 red).
The dashed lines correspond to the operation temperature of a closed-loop $^4$He refrigerator ($1.6$~K) and of a liquid $^4$He bath ($4.2$~K).}
\label{fig:4}
\end{figure}

We summarize the measurements of single and two-electron states in buried Si quantum wells in Fig.\ref{fig:4}. 
We report the charging energy ($E_C$), single electron orbital energy ($E_O$), single electron valley splitting ($E_V$), two-electrons singlet-triplet splitting from magnetospectroscopy ($E_{ST}$) and Pauli-Spin-Blockade spectroscopy ($E_{PSB}$). 
As expected, we find that the $E_V$ and $E_{ST}$ constitute the lowest excitations, followed by the $E_O$ and $E_C$. 
We report consistently high values of valley splitting, with the highest of $0.76(2)$ meV, setting a new benchmark for electrons in buried quantum wells and comparable with the best values in SiMOS. 
We find a strong correlation between orbital and valley splitting with a large proportionality coefficient of $0.22$ eV/eV, indicating that confinement can be effectively used in low-disorder heterostructures to tune and enhance the valley splitting. 
We extend our statistics on $E_{ST}$ and confirm an average of $0.24(6)$ meV across 17 lithographically identical quantum dots, over four different devices and two wafers. 

While we cannot pinpoint a specific origin for the large values reported here, new methodologies are currently being developed to identify the microscopic mechanism driving the valley splitting~\cite{woods2025statistical}.
We envision that the further characterization of these low-disorder quantum wells with such techniques and the comparison with heterostructures where the alloy disorder is intentionally enhanced~\cite{stehouwer2025engineering,thayil2025optimization,feng_enhanced_2022} will provide ultimate insight into the microscopic physics and chart a practical path toward engineering buried quantum wells into a platform that uniquely combines well-defined energy states, low electric and magnetic disorder, and compatibility with advanced semiconductor manufacturing processes.



\vspace{\baselineskip}

\bibliography{bibliography_tidy.bib}

\begin{footnotesize}
\section{Methods}
\noindent\textbf{Si/SiGe heterostructure growth.}
The $^{28}$Si/SiGe heterostructures are grown on a 100-mm n-type Si(100) substrate using an Epsilon 2000 (ASMI) reduced-pressure chemical vapor deposition reactor. 
The reactor is equipped with a $^{28}$SiH$_4$ gas cylinder (1\% dilution in H$_2$) for the growth of isotopically enriched $^{28}$Si with 800 ppm of residuals of other isotopes\cite{sabbagh_quantum_2019}. 
Starting from a Si substrate, the layer sequence of all heterostructures comprises a step-graded Si$_{(1-x)}$Ge$_x$ layer with a final Ge concentration of $x = 0.31$ achieved in four grading steps ($x = 0.07$, $0.14$, $0.21$, and $0.31$), followed by a Si$_{0.69}$Ge$_{0.31}$ strain-relaxed buffer (SRB). 
The step-graded buffer and the SRB are $\approx$3 $\upmu$m and $\approx$2.4 $\upmu$m thick, respectively. 
After the SRB, we grow a tensile-strained $^{28}$Si quantum well conformal to the substrate and thickness of $~6.9$ nm, nominally identical to \cite{degli2024low}. 
The heterostructure is then terminated with a 30 nm-thick SiGe spacer, grown under the same conditions as the virtual substrate. 
The surface of the SiGe spacer is passivated with DCS at 500~\textcelsius before exposure to air \cite{degli_esposti_wafer-scale_2022}.
We confirm the Ge concentration in the spacer and virtual substrate via secondary ions mass spectrometry (similar to Fig.S13 from ref.~\cite{paquelet_wuetz_atomic_2022}) and quantitative electron energy loss spectroscopy. 

\noindent\textbf{Device fabrication.}
The fabrication process for the quantum dot spin qubit devices comprises: reactive ion etching of mesa-trench and pre-markers; selective P-ion implantation and activation by rapid thermal annealing at 700~$^{\circ}$C; atomic layer deposition (ALD) of a 10-nm-thick Al$_2$O$_3$ gate oxide; sputtering of Al gate; selective chemical etching of the dielectric with BOE (7:1) followed by electron beam evaporation of Ti:Pt to create ohmic contacts. All patterning is performed by optical lithography at the four-inch wafer scale. Single and multi-layer quantum dot devices are fabricated on wafer coupons from the same H-FET fabrication run and share the process steps listed above. Single-layer quantum devices feature all the gates in a single evaporation of Ti:Pd (3:17~nm), followed by the deposition via ALD of a 5 nm thick AlOx layer and consequent evaporation of a global top screening gate of Ti:Pd (3:27~nm). Multi-layer quantum dot devices feature three overlapping gate metallizations with increasing Ti:Pd thicknesses (3:17~nm, 3:27~nm, 3:27~nm), each isolated by a 5 nm-thick AlOx dielectric. Finally, a 5 nm AlOx layer separates the gate stack from the micro-magnets (Ti:Co, 5:200 nm). All patterning in quantum dot devices is performed using electron-beam lithography. 

\noindent\textbf{Measurement setup and cryogenics.}
All the experiments are performed in a Blufors LD400 dilution refrigerator at a fixed temperature of $150$ mK to reduce heating effects~\cite{undseth_hotter_2023}. 
The quantum dot device is wire-bonded to a custom design PCB featuring on-chip bias Tees (RC time $\approx 100$ ms) to sum the DC and RF voltages. 
The DC voltages are supplied by a battery-powered, home-built DAC module (D5a), while the baseband control and readout pulses are generated with a series of QCM and QRM modules embedded in a QBlox Cluster. 
The readout of the two charge sensors is performed via frequency multiplexing by connecting each single-electron transistor to a tank circuit with a different resonant frequency (see Supplementary Information). 
The reflected signal is then amplified at the $4$~K stage by a Cosmic Microwave Technologies CMT-BA1 cryogenic amplifier ($\approx 30$ dB) and further amplified at room temperature ($\approx 40$~dB) to match the $1$~V dynamic range of the digitizer module. 

\noindent\textbf{Lever arm estimation.}
We extract the lever arm of each virtual gate by measuring the first-electron-addition line as a function of the mixing chamber temperature ($T_{MC}$). 
We tune the tunnel coupling to the regime where the broadening of the first addition line is limited by temperature broadening. 
In this regime, the signal ($S_i$) of the first polarization line can be expressed as:
$$
A + B + C\tanh{\alpha(V - V_0)\beta}
$$
where $A$, $B$, and $\alpha$ are fitting parameters and $\beta = \frac{e}{2 k_B T_{e}}$ with $k_B$ the Boltzmann constant and $T_e$ the electron temperature. 
For sufficiently high temperatures and good thermalization, the electron temperature can be approximated with the temperature of the  mixing chamber ($T_e \approx T_{MC}$)
We find that the condition is satisfied for $T_{MC}>0.5$ K. 
We validate this approximation by verifying that the extracted lever arm is constant in this regime, repeating the measurement at multiple temperatures between $0.5$ and $1$ K. 
We repeat the measurement for each dot, using either nearby charge sensors or nearby dots as electron reservoirs. 

\noindent\textbf{Pulsed spectroscopy.}
We perform pulsed spectroscopy of external quantum dots directly coupled to multi-electron charge sensors that act as electron reservoirs. 
We sweep the virtual plunger voltage $vP_i$ across the first electron addition line as a function of the amplitude of a superimposed square pulse of fixed frequency, typically of $f = 25$ kHz. 
We tune the tunnel coupling to be low compared to the frequency of the superimposed square wave, such that the electron tunneling to the dot is most sensitive to the presence of well-defined single-electron states.  
We record the tunneling by measuring the state of the nearby charge sensor for integration times ($t_{int}$) much longer than the period of the superimposed square wave ($t_{int} \approx 1-10$ ms $ \gg 1/f$). 

\noindent\textbf{Detuning axis pulsed spectroscopy.}
We perform detuning axis pulsed spectroscopy (DAPS) by pulsing the virtual detuning ($ve_{ij} = P_i - P_j$) between two quantum dots in a single-shot experiment and averaging the signal over multiple shots. 
We initialize the double dot in the (1,0) charge configuration by pulsing at the 'init' point deep in detuning in the (1,0) configuration and waiting for a long time, typically greater than $100$ $\mu$s. 
Next, we acquire the reference signal at the 'zero' point near the (1,0) to (0,1) charge transition by integrating the charge sensor for $5$ $\mu$s.
We then quickly pulse diabatically across the charge anticrossing at the tunneling point by sweeping the detuning $ve_{ij}$ inside the (0,1) charge state for different waiting times ($t_{wait}$). 
Finally, we pulse back to the 'zero' to acquire the measurement signal.  
We record the difference between the reference and measurement signals to filter out low-frequency charge fluctuations in the charge sensor and quantum dot device. 
We tune the tunnel coupling to be lower than the waiting time used in the measurement, thereby enhancing the contrast of resonant tunneling. 

\noindent\textbf{Magnetospectroscopy.}
 We perform magnetospectroscopy experiments in quantum dot devices cooled in a dilution refrigerator with a base temperature of $T_\textnormal{{MC}} \approx 10$~mK. 
 We use devices lithographically similar to those described in ref.~\cite{philips_universal_2022}. 
 We tune the quantum dots in the single-electron regime to isolate the 1e $\rightarrow$ 2e transition. 
 We start the magnetospectroscopy measurement from the quantum dot closest to the sensing dot, using the remaining dots as an electron reservoir. 
 We use the impedance of a nearby sensing dot to monitor the charge state of every quantum dot. 
 The impedance of the sensing dot is measured using RF reflectometry. The signal is measured by monitoring the reflected amplitude of the RF readout signal through a nearby charge sensor. 
 We fit the 1e $\rightarrow$ 2e transition as a function of the magnetic field with the relation\cite{paquelet_wuetz_atomic_2022,dodson_how_2022}:
\begin{equation}
    V_\textnormal{P} = \frac{1}{\alpha \beta_\textnormal{e}}\ln \frac{e^{\frac{1}{2} k B + \beta_\textnormal{e} E_\textnormal{{ST}}} (e^{kB} + 1) }
    {e^{kB} + e^{2kB} + e^{kB+\beta_\textnormal{e} E_\textnormal{{ST}}} + 1}
\end{equation}
where $\alpha$ is the lever arm, $V_\textnormal{P}$ is the plunger gate voltage, $E_\textnormal{{ST}}$ is the singlet-triplet energy splitting, $k = g \mu_\textnormal{B} \beta_e$, $\beta_e = 1/k_\textnormal{B} T_\textnormal{e}$, $g = 2$ is the $g$-factor in silicon, $\mu_\textnormal{B}$ is the Bohr magneton, $\textbf{B}$ is the magnetic field, $k_\textnormal{B}$ is Boltzmann's constant, and $T_\textnormal{e}$ is the electron temperature. $E_\textnormal{{ST}}$ is linked to the position of the kink ($\textbf{B}_\textnormal{{ST}}$) in the magnetospectroscopy traces by the relation $E_\textnormal{{ST}} = g \mu_\textnormal{B} \textbf{B}_\textnormal{{ST}}$.

\noindent \textbf{Valley splitting correlations.} The covariance matrix between two intervalley coupling values $\Delta_1, \Delta_2$, the covariance matrix is given in the basis $\mathbf{v}^T = [\Delta_1^R \; \Delta_1^I \; \Delta_2^R \; \Delta_2^I]$  \cite{losert_practical_2023}:

\begin{equation}
\begin{gathered}
    \mathbf{\Sigma} = \frac{1}{2}
    \begin{bmatrix}
        \sigma^2_{\Delta_1} & 0 & \left< \Delta_1, \Delta_2 \right> & 0 \\
        0 & \sigma^2_{\Delta_1} & 0 & \left< \Delta_1, \Delta_2 \right> \\
        \left< \Delta_1, \Delta_2 \right> & 0 & \sigma^2_{\Delta_2} & 0 \\
        0 & \left< \Delta_1, \Delta_2 \right> & 0 & \sigma^2_{\Delta_2}
    \end{bmatrix}
\end{gathered}
\end{equation}

\noindent where the variance and covariance matrix elements are defined:

\begin{equation}
\begin{gathered}
    \sigma_\Delta^2 = \frac{a_0^3}{16} \left( \frac{\Delta E_c}{X_w - X_s} \right)^2 \int dz |\psi_{env} (z)|^4 \bar{X} (z) (1 - \bar{X} (z)) \int dx dy |\psi(x,y)|^4 \\
    \left< \Delta_1, \Delta_2 \right> = \frac{a_0^3}{16} \left( \frac{\Delta E_c}{X_w - X_s} \right)^2 \times \\ \int dz |\psi_{env} (z)|^4 \int dx dy |\psi_1 (x,y;\vec{r}_1)|^2 |\psi_2 (x,y;\vec{r}_2)|^2
\end{gathered}
\end{equation}

\noindent where $a_0 = 0.543$ nm is the width of the conventional cubic unit cell. We will estimate the orbital wavefunctions as the ground state of the harmonic oscillator, so the variance depends on the orbital splitting $E_{OS}$ and the covariance depends on $E_{OS_1}, E_{OS_2}$ for the two dots and the distance between them $d$.

\noindent Because the $z$ part of the integral describes alloy disorder averaged over the device, we can approximate it as constant at all measurement points. We may also approximate the $y$ orbital splitting as constant by absorbing the $y$ piece of the orbital integral into this constant. Then variance can be estimated for some $E_{OS}$ by \cite{marcks2025valley}:

\begin{equation}
\begin{gathered}
    \sigma_\Delta^2 \approx \eta_{yz} \int dx |\psi(x;E_{OS})|^4
\end{gathered}
\end{equation}

\noindent We estimate the constant $\eta_{yz}$ using an averaged valley splitting and orbital splitting values across the device:

\begin{equation}
\begin{gathered}
    \eta_{yz} = \frac{\overline{\sigma}_\Delta^2}{\int dx |\psi(x;\overline{E}_{OS})|^4} \approx 4.12 \; \text{meV}^2 \cdot \text{nm}
\end{gathered}
\end{equation}

\noindent $\overline{\sigma}_\Delta^2$ is computed using maximum likelihood approximation to fit a Rayleigh distribution to the measured values of $E_{VS}$.

\noindent For $\Delta_1, \Delta_2$, we can construct the joint probability function for $\mathbf{\Sigma}$:

\begin{widetext}

\begin{equation}
\begin{gathered}
    P(E_{VS_1} < E_1, \, E_{VS_2} < E_2) = \frac{1}{2 \pi \sqrt{|\Sigma|}} \int_{|\Delta_1| < E_1 / 2} d \Delta_1 \int_{|\Delta_2| < E_2 / 2} d \Delta_2 \; \text{exp} \left( - \mathbf{v}^T \mathbf{\Sigma}^{-1} \mathbf{v} \right)
\end{gathered}
\end{equation} 

\end{widetext}

\noindent We are interested in the conditional probability $P(E_{VS_2} < E_2 \, | \, E_{VS_1} = E_1)$, which can be computed using a Monte Carlo simulation with importance sampling. For convenience, we may rewrite our covariance matrix in a blocked basis $\mathbf{v}^T = [\vec{\Delta_1} \; \vec{\Delta_1}]$:

\begin{equation}
\begin{gathered}
    \mathbf{\Sigma} = \begin{bmatrix}
        \Sigma_{11} & \Sigma_{12} \\
        \Sigma_{12} & \Sigma_{22}
    \end{bmatrix}
\end{gathered}
\end{equation}

\noindent We select some $\vec{\Delta}_1 = [\Delta_1^R \; \Delta_1^I]$ such that $|\Delta_1| = E_1 / 2$. For this known $\vec{\Delta}_1$, it can be shown \cite{losert2024strategies} that $\vec{\Delta}_2$ now follows a conditional distribution of the multivariate normal distribution $\vec{\Delta_2} \sim \mathcal{N} (\mu_{2|1}, \Sigma_{2|1})$ with conditional mean and variance defined:

\begin{equation}
\begin{gathered}
    \mu_{2|1} = \Sigma_{12} \Sigma_{11}^{-1} \vec{\Delta_1}\\
    \Sigma_{2|1} = \Sigma_{22} - \Sigma_{12} \Sigma_{11}^{-1} \Sigma_{12}
\end{gathered}
\end{equation}

\noindent Using this relationship, we may use known measurements of $E_{OS_1},E_{OS_2}$ and $E_{VS_1}$ for to approximate an expected range for $E_{VS_2}$ using importance sampling to select $\vec{\Delta_2}$.

\section{Data availability}
All data included in this work are available from the
4TU.ResearchData international data repository at \doi{10.4121/6bea90c0-1801-4aaf-a052-07c0dbbd0920}

\section{Code availability}
All the code used to derive the figures and analyze the data is included  with the data and is available at
4TU.ResearchData international data repository at \doi{10.4121/6bea90c0-1801-4aaf-a052-07c0dbbd0920}

\section{Acknowledgements}
We acknowledge technical support for device fabrication from B. van Asten, M. Fisher, and all members of the Kavli Nanolab Delft at the Van Leeuwenhoek Laboratory (VLL). 
We acknowledge helpful discussions with members of the Scappucci, Vandersypen, and Veldhorst groups at QuTech/TU Delft. 
This research was supported by the European Union's Horizon 2020 research and innovation program under the Grant Agreement No. 951852 (QLSI project) and in part by the Army Research Office (Grant No. W911NF-17-1-0274). 
The views and conclusions contained in this document are those of the authors and should not be interpreted as representing the official policies, either expressed or implied, of the Army Research Office (ARO), or the U.S. Government. 
The U.S. Government is authorized to reproduce and distribute reprints for Government purposes, notwithstanding any copyright notation herein.
This research was sponsored in part by The Netherlands Ministry of Defence under Awards No. QuBits R23/009. 
The views, conclusions, and recommendations contained in this document are those of the authors and are not necessarily endorsed nor should they be interpreted as representing the official policies, either expressed or implied, of The Netherlands Ministry of Defence. 
The Netherlands Ministry of Defence is authorized to reproduce and distribute reprints for Government purposes notwithstanding any copyright notation herein.

\section{Authors contributions}
\noindent 

D.D.E. fabricated the quantum dot spin qubit devices and performed the experiment with input from A.E. and D.C. 
D.D.E. designed and built the experimental setups with the support from A.E. and D.C. 
D.D.E. analyzed the data with input from all the authors. 
E.C.B. and M.F. performed the valley-splitting simulations. 
D.D.E. and G.S. conceived the project. 
D.D.E., E.C.B, M.F., and G.S. wrote the manuscript with input from all authors.
M.F. and G.S. supervised the project. 

\section{Competing Interests}
The authors declare no competing interests.

\section{Additional information}
\noindent \textbf{Supplementary Information} Supplementary Figures~1--12 and Supplementary Tables~1--2.

\noindent \textbf{Correspondence and request for materials} should be addressed to G.S. and D.D.E.
\end{footnotesize}

\end{document}